\documentclass{ifacconf}

\usepackage{graphicx}      
\usepackage{natbib}        

\usepackage{inputenc}
\usepackage{textcomp}

\usepackage{amsfonts}

\usepackage{amsmath}
\usepackage[version=4]{mhchem}
\usepackage{longtable,tabularx}

\usepackage{epstopdf}
\usepackage{cases}
\usepackage{color}

\usepackage{overpic}

\usepackage{array}
\usepackage{multirow}
\usepackage{bigdelim}
\usepackage{amssymb}

\usepackage{tikz}
\usetikzlibrary{positioning,arrows.meta,calc, shapes, arrows}

\usepackage{ctable}
\usepackage{mdwtab}

\usepackage{bm}

\DeclareMathOperator{\sign}{sgn}

\newcommand{\R}{\mathbb{R}} 
\newcommand{\N}{\mathbb{N}} 

\renewcommand{\S}[2]{  \ensuremath{S_{[#1,#2]}}}

\DeclareMathOperator*{\argmin}{arg\,min}
\DeclareMathOperator*{\argmax}{arg\,max}

\usepackage{accents}

\newtheorem{remark}{Remark}

\newtheorem{problem}{Problem}

\usepackage{color}

\begin{document}
\begin{frontmatter}

\title{From open-loop representations to closed-loop feedback implementations in differential games: A numerical case study\thanksref{footnoteinfo}} 

\thanks[footnoteinfo]{P. Braun, T. Molloy and I. Shames are supported by the United States Air Force Office of Scientific Research
under Grant No. FA2386-24-1-4014.}

\author[First]{Philipp Braun} 
\author[Fourth]{Timothy L. Molloy} 
\author[Second]{Gal Barkai} 
\author[Third]{Iman Shames}

\address[First]{Australian National University, Canberra, Australia (e-mail: philipp.braun@anu.edu.au)}
\address[Fourth]{Monash University, Melbourne, Australia (e-mail: timothy.molloy@monash.edu)}
\address[Second]{University of Lorraine, France, (e-mail: gal.barkai@univ-lorraine.fr)}
\address[Third]{University of
Melbourne, Melbourne, Australia, (email: iman.shames@unimelb.edu.au)}

\begin{abstract}                
Solutions to pursuit-evasion and surveillance-evasion differential games are typically computed and expressed using open-loop representations, with the synthesis of feedback strategies significantly less common. We propose a numerical scheme for obtaining feedback strategies for the recently introduced prying-pedestrian surveillance-evasion differential game. The scheme involves computing feedback strategies as input-output maps approximated via neural networks trained using data obtained from open-loop representations of solutions. Simulations show the effectiveness of neural networks trained with an appropriate learning-loss function. Since optimal feedback strategies are discontinuous, as a second contribution, the potential loss/gain of individual players is subsequently studied for players using sample-and-hold feedback compared to continuous-time feedback.
\vspace*{-0.2cm}
\end{abstract}


\end{frontmatter}

\section{Introduction}

Since the pioneering work of  \cite{von1944theory}, game theory has become an established research topic, and has seen seminal contributions by \cite{Isaacs65} and others \citep{Merz1971,Lewin2012,Basar1999} in its differential (or dynamic) form.
Solutions of nowadays classical differential games have received perhaps the most attention in control.
Classical differential games include pursuit-evasion games such as the homicidal chauffeur \citep{Isaacs65, Weintraub2020,Merz1971} and suicidal pedestrian \citep{Exarchos2015,Exarchos2014,Exarchos2016}; surveillance-evasion games as discussed in \citep{Dobbie1966,Taylor1970,Lewin1975}; and collision avoidance games \citep{Merz1973,Miloh1976,Olsder1978}.

\vspace{-0.1cm}

Solutions of classical differential games (as outlined in detail in \citep{Basar1999,Lewin2012}, for example) are typically developed using open-loop representations, not feedback laws.
Feedback laws can, in principle, be obtained from these open-loop representations by exploiting equivalence relationships between costates and the gradient of the value function when it is continuously differentiable   (cf.~\citep[Thm.~8.2]{Basar1999}).
However, in many classical differential games, the value function is Lipschitz continuous, but not continuously differentiable on the entire game set.
In particular, the game set may contain lower dimensional surfaces (i.e., set of measure zero) where the value function is not continuously differentiable.
Whilst viscosity solutions of the Isaacs equation provide a means of addressing such differentiability concerns (see \citep[Eq. (8.6)]{Basar1999}, \citep[Sec. 8.2.1 \& 8.2.2]{Basar1999}, and \citep[Ch. VIII]{bardi1997optimal}, for example), the optimal feedback laws (corresponding to pure Nash equilibria) in these areas of the game set may be not uniquely defined. 
(See \cite[Ch. 9]{Lewin2012} for a discussion of the topography of the optimal value function, for example.)

\vspace{-0.1cm}

Non-uniqueness of the optimal feedback law can lead to dilemmas where one player may benefit from making a decision that is not aligned with the other player's decision.
This problem is for example discussed in \cite[Ch. 6]{Isaacs65}.
In particular, in \cite[Ch. 6.4]{Isaacs65} a perpetual dilemma is shown to arise in a wall-pursuit game for certain initial conditions in which both players are under the dilemma of going up or down, depending on the knowledge of the other players strategy.
The perpetual dilemma of the wall-pursuit game is discussed in detail in \cite{9661291}, where the potential loss of players is characterized through the rate of loss for hold times of the optimal feedback strategies, i.e., the change in the value functions is analyzed if initially a sample-and-hold implementation of the optimal feedback law instead of a continuous implementation of the feedback law is used on the surface causing the perpetual dilemma.
The perpetual dilemma of non-unique optimal controls and the potential gain/loss for individual players is also mentioned in \cite[Rem. 1]{Exarchos2015} in the context of the suicidal-pedestrian game.

\vspace{-0.1cm}

The work discussed in this paper is inspired by that of \cite{9661291}, and in particular the potential loss and gain in the optimal value function captured through the loss in \cite[Sec. V]{9661291} during hold times  (\cite[Def. 2]{9661291}).
Here, for the prying-pedestrian surveillance-evasion game introduced in \cite{prying_pedestrian}, we investigate the potential gain/loss in terms of the value of the game when continuous-time feedback laws are replaced by sample-and-hold feedback implementations.
This setup is related to Friedman's theory as developed in \citep{friedman2013differential} (and outlined in \cite[Ch. 3]{bardi1997optimal}) where players only make decisions at discrete sampling times but the game dynamics evolve in continuous time.
For the numerical performance analysis of 
approximate feedback laws in the prying-pedestrian differential game, 
we use open-loop solution representations obtained in \cite{prying_pedestrian} to learn the optimal value function and feedback laws using a neural network. 
These learned feedback laws enable us to investigate the potential loss/gain in performance of each player when practical sample-and-hold implementations of the feedback laws are used (instead of idealized continuous feedback). 
The results show that players can gain an advantage over their opponent in neighborhoods of dispersal surfaces along the $y$-axis by sampling at a comparatively higher rate (or lesser period) due to the discontinuity of optimal feedback laws at these surfaces.

\vspace{-0.1cm}

The paper is structured as follows. In Section \ref{sec:setting}, the prying-pedestrian differential game and an implicit definition of optimal feedback laws relying on the knowledge of the derivatives of the optimal value function are recalled.
Section \ref{sec:neural_network_approximation} illustrates how open-loop representations of solutions to the game can be used to train a neural network to obtain feedback laws.
Since the optimal feedback laws are discontinuous on the $y$-axis, Section \ref{sec:sample_and_hold} investigates the impact of sample-and-hold control laws in terms of the potential loss/gain of individual players.
The paper concludes with final remarks and future work in Section \ref{sec:conclusions}.

\vspace{-0.1cm}

Throughout the paper, the real numbers and the positive real numbers are denoted by $\R$ and $\R_{>0}$, respectively. Similarly, for $n\in \N$, where $\N$ denotes the natural numbers, $\R^n$ denotes the Euclidean space of dimension $n$. The $1$-norm and $2$-norm in $\R^n$ are denoted by $|\cdot|_1$ and $|\cdot|_2$, respectively.

\section{Setting and problem formulation} \label{sec:setting}

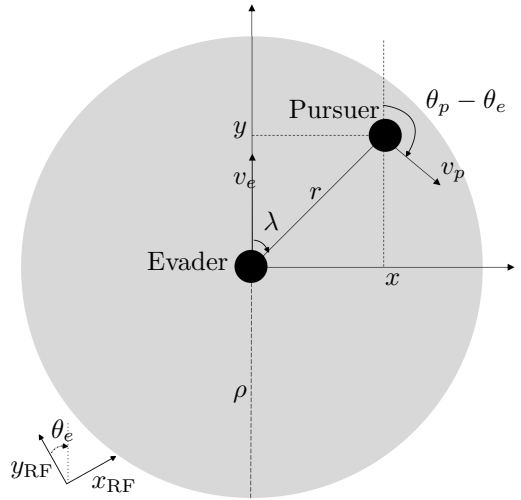
\begin{figure}[t!]
    \centering
    \begin{overpic}[width = 0.75\columnwidth]{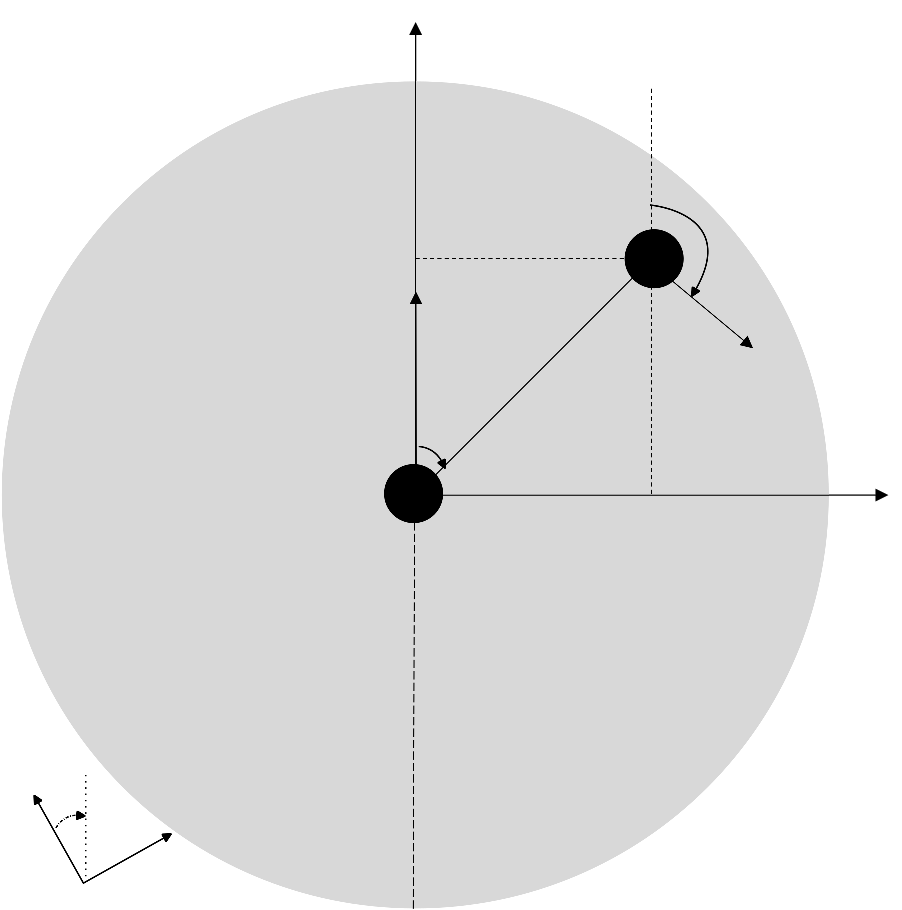}
    \put(25,45){Evader}
    \put(53,75){Pursuer}
    \put(42,20){$\rho$}
    \put(42,72){$y$}
    \put(42,62){$v_e$}
    \put(48,53){$\lambda$}
    \put(57,59){$r$}
    \put(72,42){$x$}
    \put(83,64){$v_p$}
    \put(80,77){$\theta_p-\theta_e$}
    \put(14,2){$x_{\text{RF}}$}
    \put(-2,5){$y_{\text{RF}}$}
    \put(6,12){$\theta_e$}
    \end{overpic}
    \caption{Coordinate system attached to the evader.}
    \label{fig:fig1}
\end{figure}

Following the presentation in \cite{prying_pedestrian}, we consider the prying-pedestrian surveillance-evasion game of degree consisting of an agile but slow pursuer and a fast but less maneuverable evader.
The evader and the pursuer are moving in the two-dimensional Euclidean plane.
The evader has unicycle (or Dubins car) kinematics given by
\begin{align}
   \label{eq:cartesianDynamics}
\dot{\xi}_e = \left[ \begin{array}{c}
     \dot{x}_e  \\
     \dot{y}_e \\
     \dot{\theta}_e
\end{array} \right] = 
\left[\begin{array}{cc}
      v_e \sin \theta_e(t)\\ 
      v_e \cos \theta_e(t)\\
    \omega_e u_e(t).
\end{array} \right]
\end{align}
with state $\xi_e = [x_e,y_e,\theta_e]^\top \in \mathbb{R}^3$ capturing the position and the heading angle, input  $u_e\in[-1,1]$ denoting the normalized turning rate and parameters $v_e, \omega_e \in \R_{>0}$ defining the speed and the maximal turning rate.
The (agile) pursuer has kinematics given by
\begin{align}
   \label{eq:cartesianDynamicsPursuer}
\dot{\xi}_p = \left[ \begin{array}{c}
   \dot{x}_p(t) \\
   \dot{y}_p(t)
\end{array} \right] =
\left[
\begin{array}{c}
   v_p \sin \theta_p(t)\\
   v_p \cos \theta_p(t)
\end{array}
\right]
   \end{align}
   with state $\xi_p = [x_p,y_p]^\top \in \mathbb{R}^2$, input $\theta_p \in (-\pi,\pi]$ and parameter $v_p\in \R_{>0}$ denoting the speed.
The pursuer is agile in the sense that it is capable of instantaneously changing its direction of movement through $\theta_p(t)$, while the evader is faster than the pursuer by assumption, i.e., $v_e>v_p$.
Using the coordinate transformation 
\begin{align}
    \xi &=\left[\begin{array}{c}
         x \\
         y
 \end{array} \right] 
 = \left[\begin{array}{rr}
     \cos\theta_e & -\sin\theta_e  \\
     \sin\theta_e & \cos\theta_e 
 \end{array} \right] (\xi_p -\xi_e) \in \mathbb{R}^2, \label{eq:coordinate_transformation_xi}
\end{align}
we can change the (inertial) coordinates in the global reference frame to an evader-centric coordinate system,
(i.e., the origin is at $\xi_e$); its $y$-axis aligned with the evader's heading $\theta_e$; and, its $x$-axis orientated at an angle of $\pi/2$ radians clockwise from the positive $y$-axis (as shown in Fig.~\ref{fig:fig1}).
With the additional input transformation
\begin{align}
u_p(t) = \theta_p(t) - \theta_e(t) \in (-\pi,\pi],  \label{eq:input_u_p}
\end{align}
defining the new input of the pursuer, the evader-centric dynamics can be written as
\begin{align}
    \label{eq:dynamics}
    \dot{\xi}(t)
    &= f(\xi(t), u_e(t), u_p(t)), \\
    f(\xi, u_e, u_p)
    &=
    \begin{bmatrix}
        -\omega_e y u_e + v_p \sin u_p\\
        \omega_e x u_e - v_e + v_p \cos u_p
    \end{bmatrix}. \label{eq:f_dynamics}
\end{align} 
A derivation of the dynamics \eqref{eq:dynamics} can be found in \cite{prying_pedestrian}.

Solutions of \eqref{eq:dynamics} with respect to initial condition $\xi_0\in \R^2$, in forward time $t \in \R_{\geq 0}$ or  backward time $\tau = -t $, for inputs $u_e,u_{e_\tau}:\R_{\geq 0} \rightarrow [-1,1]$, $u_p,u_{p_\tau}:\R_{\geq 0} \rightarrow (-\pi,\pi]$, are denoted by
\begin{align}
\left[ \begin{smallmatrix}
        x(\cdot) \\
        y(\cdot)
    \end{smallmatrix} \right] &= 
    \xi(\cdot) = \xi_t(\cdot;t_0,\xi_0,u_e(\cdot),u_p(\cdot)) \\ 
    \left[ \begin{smallmatrix}
        x_\tau(\cdot) \\
        y_\tau(\cdot)
    \end{smallmatrix} \right] &=
    \xi_\tau(\cdot) = \xi_\tau(\cdot;\tau_0,\xi_0,u_{e_\tau}(\cdot),u_{p_\tau}(\cdot)), \label{eq:solution_notation}
\end{align}
respectively. Throughout this paper we assume that $u_{e_\tau}(\cdot)$, $u_{p_\tau}(\cdot)$ are measurable functions.

With these definitions, the prying-pedestrian surveillance-evasion game of degree can be summarized as follows.

\begin{problem}[Game of Degree] \label{prob:game_of_degree}
Consider the dynamics \eqref{eq:dynamics} with state $\xi\in \R^2$, inputs $u_e\in [-1,1]$ and $u_p\in \R$,
defined through parameters $\omega_e,v_e,v_p, \rho\in \R_{>0}$ with $v_p<v_e$ and game set defined as 
\begin{align}
 \mathcal{S}=\{\xi\in \R^2| \ |\xi|_2 \leq \rho\}. \label{eq:S}
\end{align}
The game of degree with surveillance radius $\rho$ is defined through the optimization problem\footnote{For simplicity of notation the condition  
$\frac{d}{dt}|\xi(T)|_2^2 >0$
is used instead of the condition $f(\xi(T),u_e(T),u_p(T))^\top \xi(T) > 0$, which does not rely on the differentiability of $\xi(\cdot)$ at time $T$. }
\begin{align}
V(\xi_0) = &\min_{u_e:[0,T]\rightarrow[-1,1]} \max_{u_p:[0,T]\rightarrow\R}  \int_0^T 1 \, \mathrm{d}t \label{eq:problem_constraints} \\
\begin{split}
\text{subject to } \quad & \dot{\xi}(t) = f(\xi(t), u_e(t), u_p(t)), \quad \xi(0) = \xi_0, \\
& |\xi(0)|_2 \leq   \rho, \quad |\xi(T)|_2 = \rho, \quad \tfrac{d}{dt} |\xi(T)|_2^2 >0. 
\end{split}  \nonumber
\end{align}
\end{problem}

The approach described in \cite{prying_pedestrian} provides open-loop representations of solutions to Problem \ref{prob:game_of_degree} with optimal inputs
$u_{e_\tau}:\R_{\geq 0} \rightarrow [-1,1]$, $u_{p_\tau}:\R_{\geq 0} \rightarrow \R$ derived as functions of time $\tau$.
However, as it is common in the differential game literature, the approach does not lead to explicit expressions of optimal feedback laws $u_e^*: \mathcal{S} \rightarrow  [-1,1]$ and  $u_p^*: \mathcal{S} \rightarrow \R$, i.e., functions of the state $\xi$ instead of functions of time $\tau$ (or $t$)\footnote{While the approach used in \cite{prying_pedestrian} in general does not return a feedback law, a feedback law $u_{e}(\xi)$ can be easily deduced from the representation in \cite{prying_pedestrian}. However, a feedback law $u_{p}(\xi)$ of the pursuer is not readily available.}. 

Under the assumption that the optimal value function $V:\mathcal{S} \rightarrow \R_{\geq 0}$ is continuously differentiable, the Hamiltonian function can be defined as
\begin{align}
\label{eq:Ham}
H(\xi,\nabla V(\xi),u_e, u_p) =  \nabla V(\xi)^{\top} f(\xi, u_e, u_p) + 1,
\end{align}
which satisfies Pontryagin's Principle
\begin{align}
\label{eq:Ham_cond1}
\min_{u_e \in [-1,1]} H(\xi,\nabla V(\xi),u_e, u_p^*) &\leq H(\xi,\nabla V(\xi),u_e^*, u_p^*)  \\
& \leq \max_{u_p \in \R} H(\xi,\nabla V(\xi),u_e^*, u_p), \nonumber
\end{align}
see also \cite[Ch. 8]{Basar1999}.
Thus, if the optimal value function implicitly defined in \eqref{eq:problem_constraints} is known and Lipschitz continuous, then we can calculate the gradient $\nabla V(\xi)$ for almost all $\xi \in \mathcal{S}$, which allows us to define optimal feedback laws
\begin{align}
\begin{split}
    u_e^*(\xi) &= \argmin_{u_e \in [-1,1]}  
    \left[   (\nabla_y V(\xi) x -\nabla_x V(\xi) y)  \omega_e u_e \right] \\
     & = \sign(\nabla V_x(\xi) y-\nabla V_y(\xi)  x) 
     \end{split}\label{eq:opt_feedback_law_e} \\
     \begin{split}
    u_p^*(\xi) &=     \argmax_{u_p(\xi) \in \R} \left[ (\nabla V_x(\xi)  \sin u_p  + \nabla V_y(\xi) \cos u_p) v_p      \right] \\
    &= \arctan_2\big(\nabla V_x(\xi),\nabla V_y(\xi)\big)
    \end{split}\label{eq:opt_feedback_law_p}
\end{align}
for almost all $\xi \in \mathcal{S}$ (and $\arctan_2$ is defined as in \cite[Ch. 1]{kellett2023introduction}).
To extend the definitions of $u_e^*$ and $u_p^*$ to the set $\mathcal{S}$, we define the set of measure zero 
\begin{align}
    \mathcal{N}=\{\xi \in \mathcal{S}| \ \nabla V(\xi) \text{ does not exist} \}
\end{align}
and define the feedback laws 
\begin{align}
     u_e^*: \mathcal{S} \rightrightarrows  [-1,1], \qquad  u_p^*: \mathcal{S} \rightrightarrows \R \label{eq:set_valued_feedback_laws}
\end{align}
as set-valued maps, using 
\begin{align*}
u_e^*(\xi) \hspace{-0.06cm} &  \in  \hspace{-0.06cm}  \left\{ \hspace{-0.025cm} \limsup_{i\rightarrow \infty} u_e^*(\xi_i) | \{\xi_i\}_{i\in \N}\subset \mathcal{S}\backslash \mathcal{N},  \lim_{i\rightarrow \infty} \xi_i \hspace{-0.03cm}=\hspace{-0.025cm} \xi \right\} \\
u_p^*(\xi) \hspace{-0.06cm} & \in \hspace{-0.06cm}   \left\{ \hspace{-0.04cm} \limsup_{i\rightarrow \infty} u_p^*(\xi_i) | \{\xi_i\}_{i\in \N}\subset \mathcal{S}\backslash \mathcal{N},  \lim_{i\rightarrow \infty} \xi_i \hspace{-0.03cm}=\hspace{-0.03cm} \xi \hspace{-0.03cm} \right\}\hspace{-0.05cm}.
\end{align*}
It follows from the derivations and illustrations in \cite{prying_pedestrian} that $V(\cdot)$ is continuously differentiable for almost all $\xi \in \mathcal{S}$.

\begin{remark}
The representations
\eqref{eq:opt_feedback_law_e} and
\eqref{eq:opt_feedback_law_p} are a direct consequence of \eqref{eq:Ham} and \eqref{eq:Ham_cond1}, which additionally shows that $u_e^*$ and $u_p^*$ can be calculated independently (if $V(\xi)$ is continuously differentiable). Accordingly, the min-max problem formulation \eqref{eq:problem_constraints} can be equivalently written in terms of a max-min optimal control problem 
\citep[Remark 8.1]{Basar1999}.
\end{remark}

In this paper, we will use the open-loop representations of solutions obtained via the methods discussed in \cite{prying_pedestrian} to approximate the optimal value function $V(\cdot)$ and its derivatives so as to define optimal feedback laws \eqref{eq:set_valued_feedback_laws}. 
In addition, we will investigate the performance of the feedback laws when compared with the optimal (open-loop) trajectories and we will discuss the potential impact of sample-and-hold feedback strategies on the performance and performance loss of both players.

\section{Feedback law approximations of the game of degree} \label{sec:neural_network_approximation}

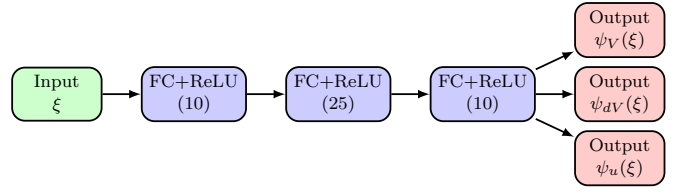
\begin{figure}[t!]
\resizebox{1\columnwidth}{!}{
\begin{tikzpicture}[
  >=latex,
  node distance=12mm and 6mm,
  layer/.style={
    draw,
    rounded corners=2mm,
    minimum width=14mm,
    minimum height=8mm,
    align=center,
    thick,
    font=\footnotesize
  },
  fc/.style={layer, fill=blue!20},
  input/.style={layer, fill=green!20},
  output/.style={layer, fill=red!20}
]

\node[input] (in) {Input \\ $\xi$};
\node[fc, right=of in] (fc1) {FC+ReLU \\ (10)};
\node[fc, right=of fc1] (fc2) {FC+ReLU \\ (25)};
\node[fc, right=of fc2] (fc3) {FC+ReLU \\ (10)};

\node[output, right=of fc3, yshift=10mm] (out1) {Output \\ $\psi_V(\xi)$};
\node[output, right=of fc3] (out2) {Output \\ $\psi_{dV}(\xi)$};
\node[output, right=of fc3, yshift=-10mm] (out3) {Output \\  $\psi_u(\xi)$};

\draw[->, thick] (in) -- (fc1);
\draw[->, thick] (fc1) -- (fc2);
\draw[->, thick] (fc2) -- (fc3);

\draw[->, thick] (fc3) -- (out1);
\draw[->, thick] (fc3) -- (out2);
\draw[->, thick] (fc3) -- (out3);

\end{tikzpicture}
}
\caption{Neural network used to obtain approximations of the optimal value function $V(\xi)$, the gradient $\nabla V(\xi)$ and the optimal inputs $(u_e(\xi),u_p(\xi))$.}
\label{fig:neural_network}
\end{figure}

In this section, we use open-loop data to compute feedback laws \eqref{eq:set_valued_feedback_laws} as well as $V(\cdot)$ and $\nabla V(\cdot)$.
We use a neural network with the network structure shown in Fig. \ref{fig:neural_network} and defining functions
\begin{align}
    &\psi_V:\mathcal{S} \rightarrow \R, \  \psi_{dV}: \mathcal{S} \rightarrow \R^2, \  \psi_u: \mathcal{S} \rightarrow [-1,1] \times \R,
\end{align}
where $\psi_V(\cdot)$ denotes an approximation of the optimal value function $V(\cdot)$, $\psi_{dV}(\cdot)$ denotes an approximation of the gradient (or costates) $\nabla V(\cdot)$ and $\psi_u(\cdot)$ denotes an approximation of the optimal feedback laws $(u_e^*(\cdot),u_p^*(\cdot))$%
\footnote{%
It would be sufficient to only learn the function $\xi\mapsto \psi_u(\xi)$. The additional components $\psi_V(\cdot)$ and $\psi_{dV}(\cdot)$ are included to obtain a complete picture of the game of degree in Problem \ref{prob:game_of_degree}.}.
The three functions are summarized as 
\begin{align}
        \psi(\cdot)=\left[\begin{smallmatrix}
        \psi_V(\cdot) \\
        \psi_{dV}(\cdot) \\
        \psi_u(\cdot)
    \end{smallmatrix}\right],
\end{align}
corresponding to the output of the neural network.

To train the network, the following loss function is used
\begin{align}
    L(\xi) &=  \kappa_{10}( \psi_V (\xi) - V(\xi)) + \kappa_{10}(\psi_{dV} (\xi) - \nabla V(\xi)) \notag \\
    & \quad + \kappa_{10} \left( \psi_u (\xi) -  \left[ \begin{smallmatrix}
        u_e^*(\xi) \\
        u_p^*(\xi)
    \end{smallmatrix} \right] \right)  + \kappa_{10} \left( \nabla \psi_V (\xi) - \psi_2(\xi) \right) \notag \\
    & \quad + \kappa_{10} \left( \left[ \begin{smallmatrix}
        \sign(\psi_{dV,1}(\xi)\xi_2 - \psi_{dV,2}\xi_1)- \psi_{u,1}(\xi) \\
        \arctan_2(\psi_{dV,1}(\xi),\psi_{dV,2}(\xi)) - \psi_{u,2}(\xi)
    \end{smallmatrix}  \right] \right)
     \label{eq:loss_function}
\end{align}
where
    $\kappa_{10}(x) = 10 |x|_1 + |x|_2^2$
denotes a linear combination of the 1-norm and the 2-norm.

The data for the training is generated by simulating open-loop representations of solutions obtained through the method described in \cite{prying_pedestrian} in backwards time and for parameters selected as $\rho=1$, $v_e=1.5$, $v_p=1$ and $w_e=1$. The method returns data of the form
\begin{align}
    (\xi_1,\xi_2,\nabla_x V(\xi),\nabla_y V(\xi),V(\xi)),
\end{align}
which in combination with \eqref{eq:opt_feedback_law_e} and \eqref{eq:opt_feedback_law_p} can be used in the loss function \eqref{eq:loss_function}.

\begin{figure}[t!]
    \centering
    \begin{overpic}[width = 0.8\columnwidth]{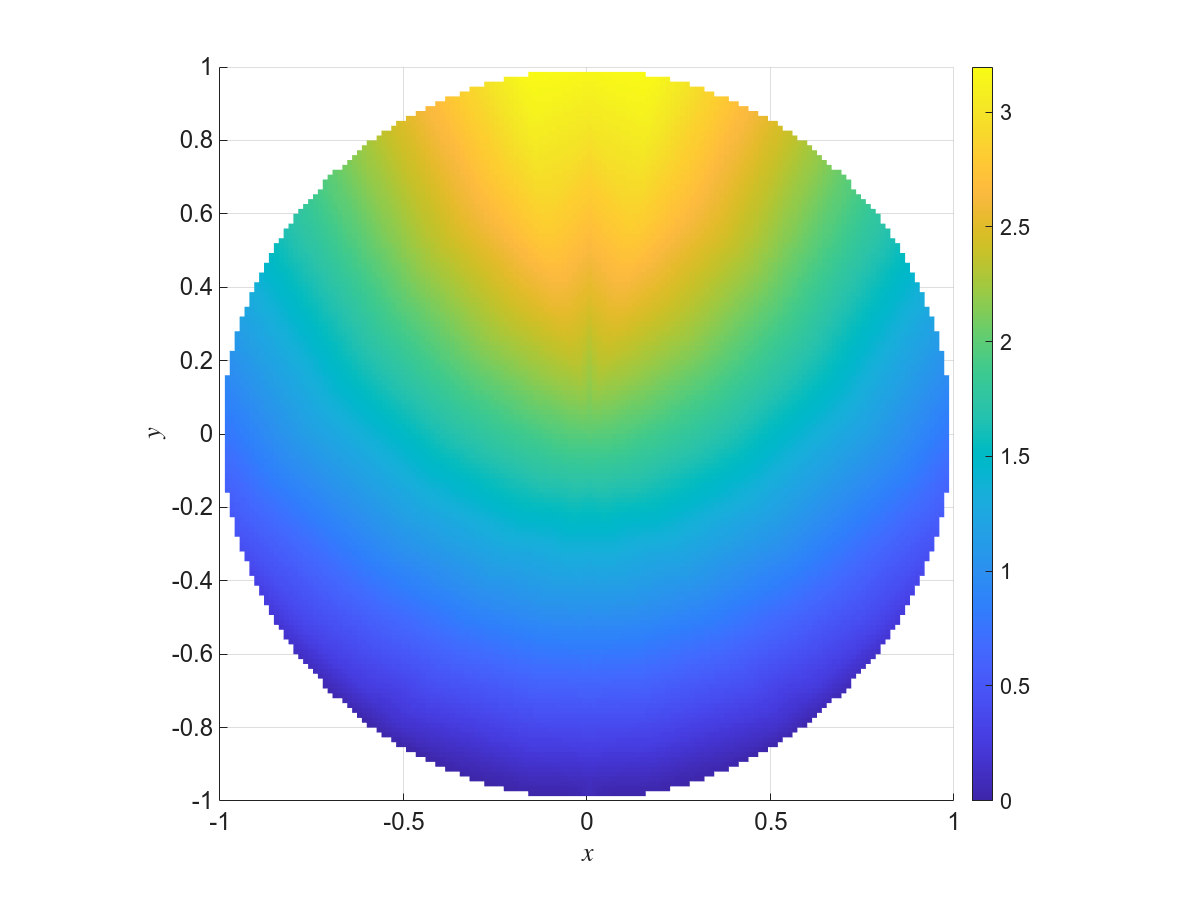}
    \end{overpic}
    \caption{Approximation of the optimal value function $\psi_{V}(\xi)$.}
    \label{fig:V}
\end{figure}

\begin{figure}[t!]
    \centering
    \begin{overpic}[width = 0.48\columnwidth]{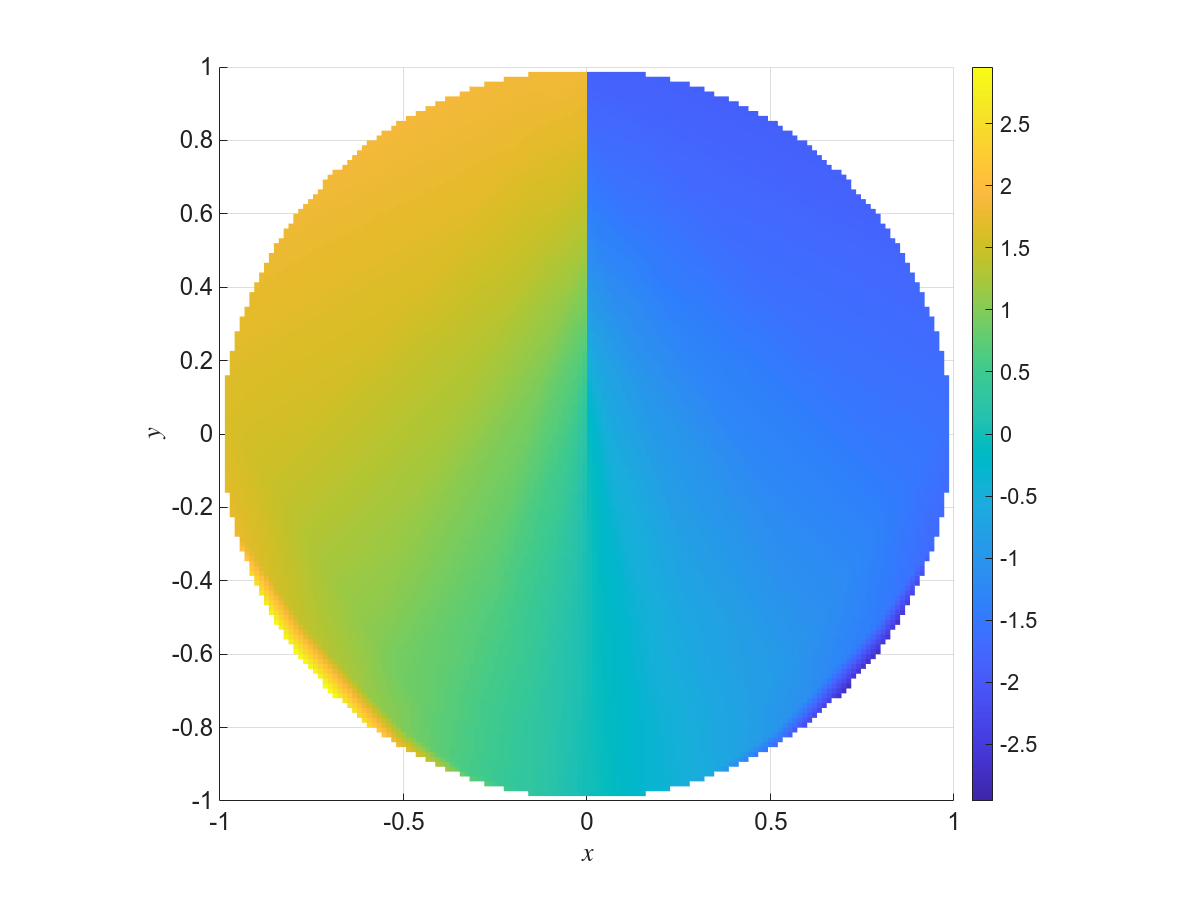}
    \end{overpic}
    \begin{overpic}[width = 0.48\columnwidth]{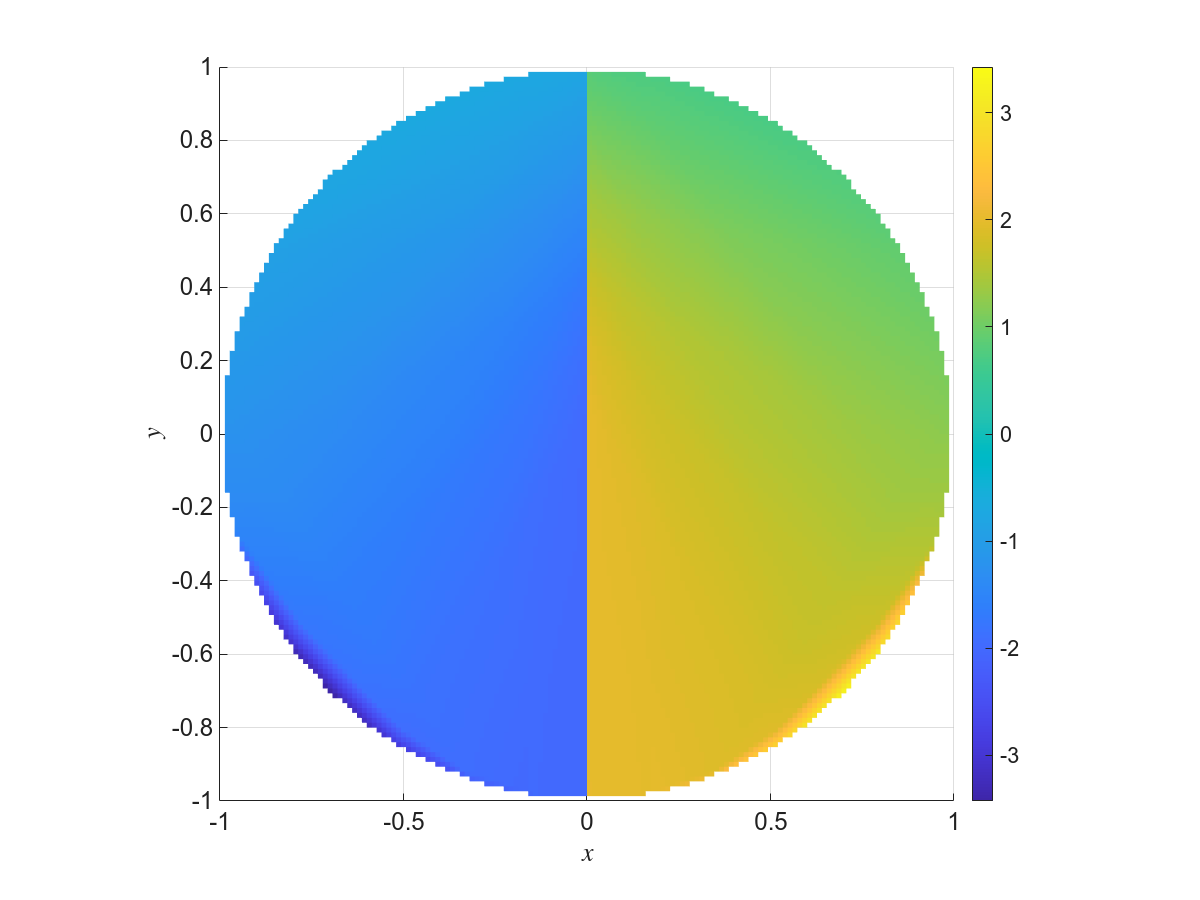}
    \end{overpic}
    \caption{Approximation of the gradient of $V$ in terms of $\psi_{dV,1}(\xi)$ on the left and $\psi_{dV,2}(\xi)$ on the right.}
    \label{fig:grad_V}
\end{figure}

\begin{figure}[t!]
    \centering
    \begin{overpic}[width = 0.48\columnwidth]{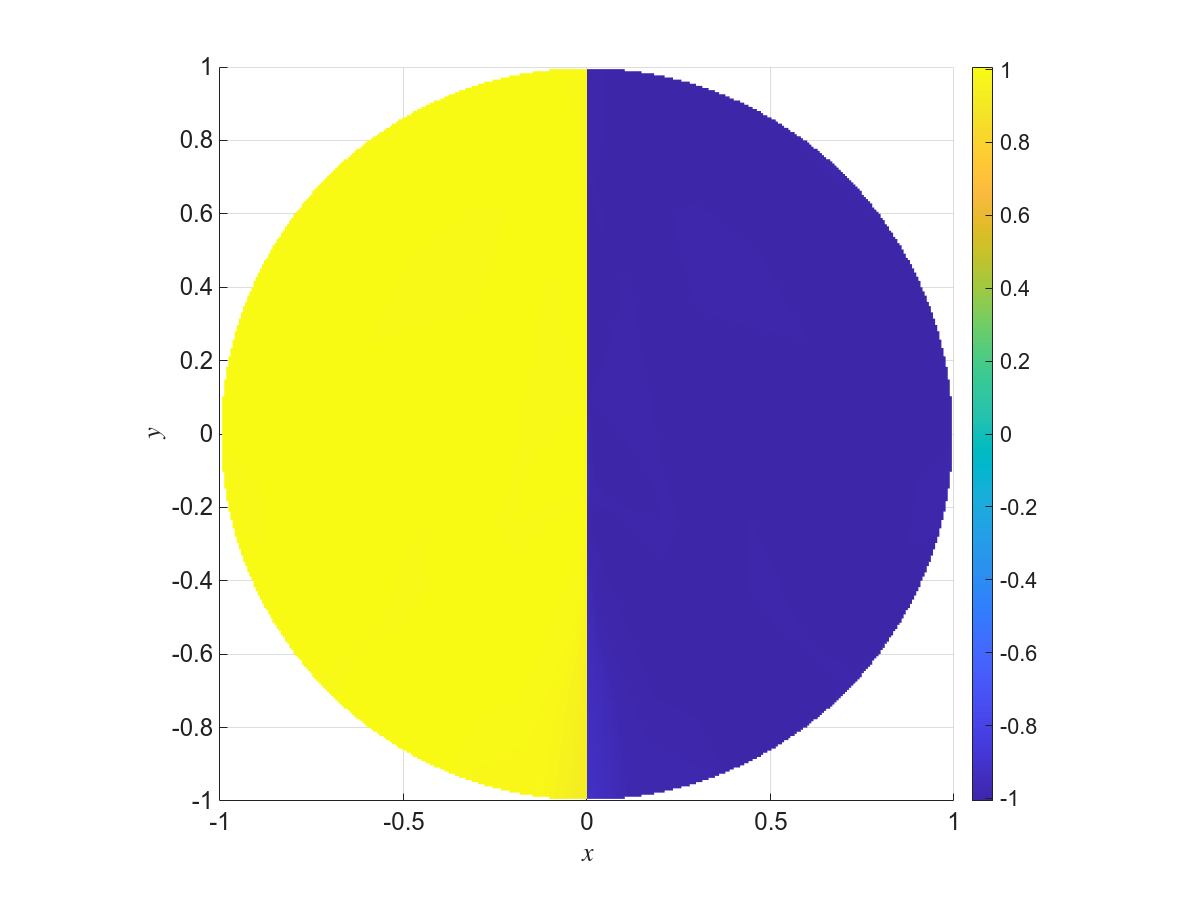}
    \end{overpic}
    \begin{overpic}[width = 0.48\columnwidth]{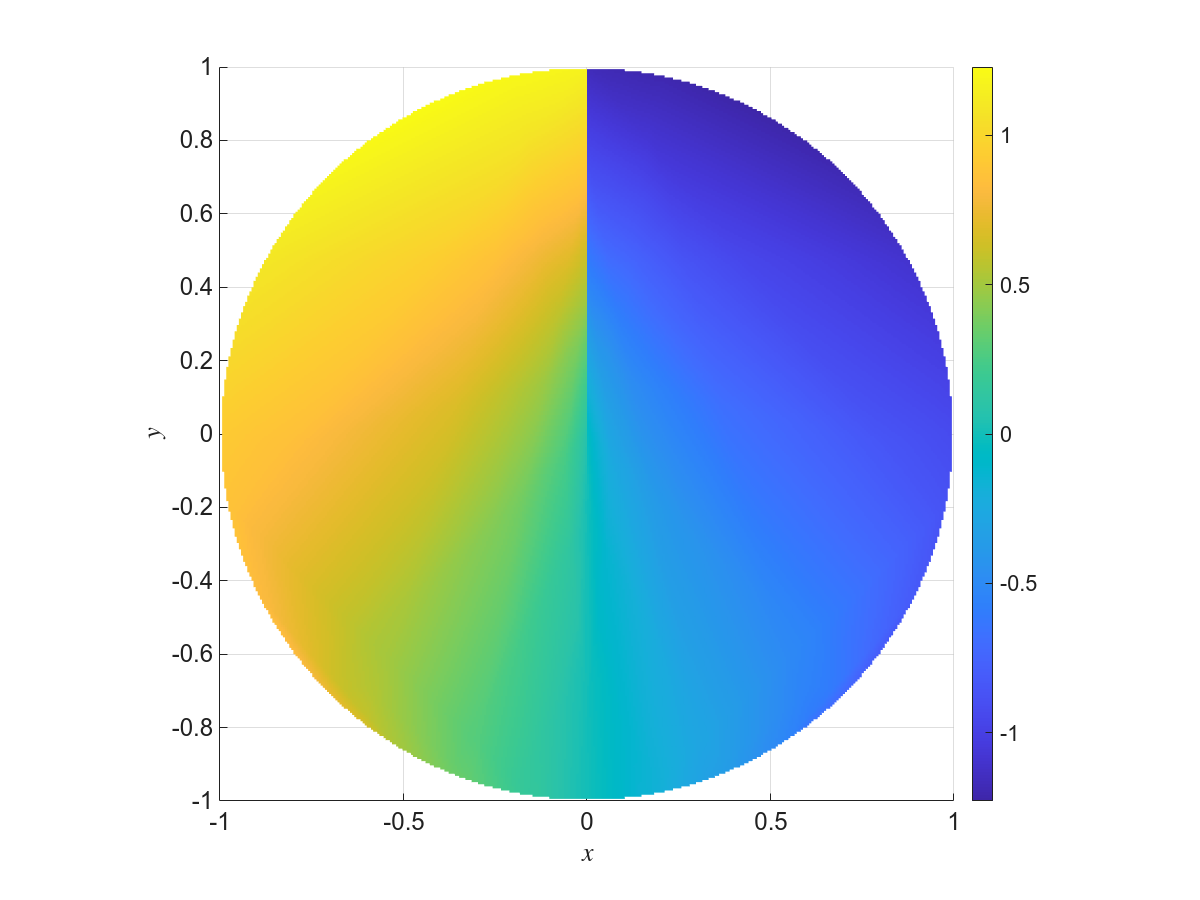}
    \end{overpic}
    \caption{Feedback law $\psi_{u,1}(\xi)$ (left) and $\psi_{u,2}(\xi)$ (right).}
    \label{fig:u_e_i_p}
\end{figure}

\begin{figure}[htb!]
    \centering
    \begin{overpic}[width = 0.7\columnwidth]{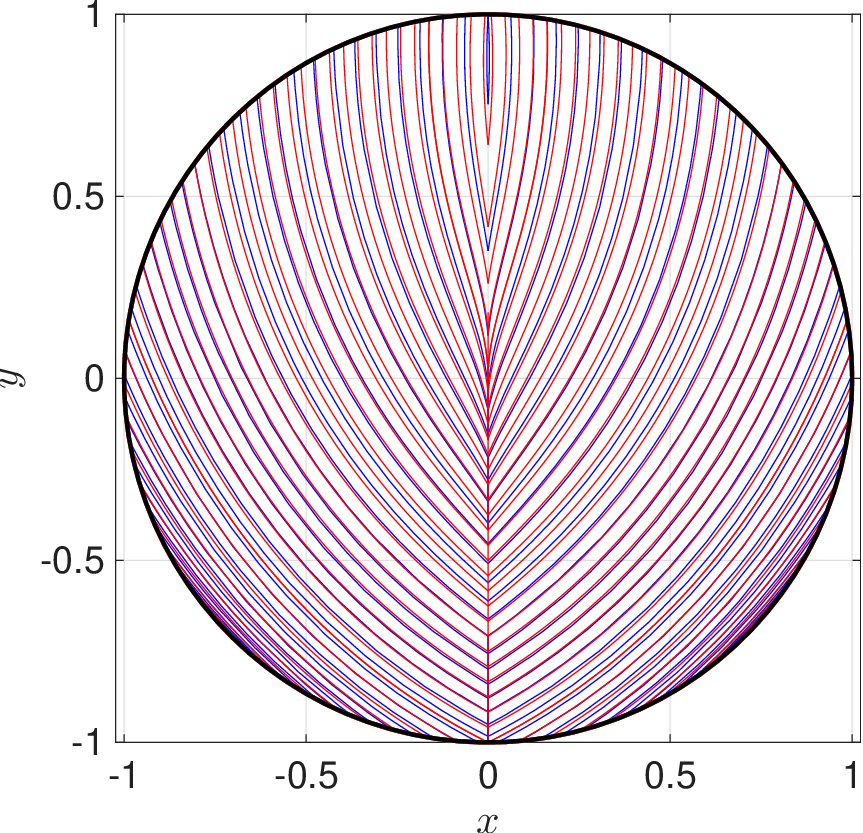}
    \end{overpic}
    \caption{Open-loop solution representations (red) obtained via the method described in \cite{prying_pedestrian} by simulating the dynamics backwards in time compared with solutions defined through the feedback law $\psi_e(\xi)$.}
    \label{fig:sol_comparison}
\end{figure}

Fig. \ref{fig:V} shows contour lines of the approximation of the optimal value function $\psi_V(\cdot)$, Fig. \ref{fig:grad_V} shows the gradient of the approximation of the optimal value function $\psi_{dV}(\cdot)$ and Fig. \ref{fig:u_e_i_p} shows the feedback laws $\psi_{u,1}(\xi)$ and $\psi_{u,2}(\xi)$ given by the neural network. 
Fig. \ref{fig:sol_comparison} shows open-loop representations of solutions in red and solutions obtained through the feedback law $\psi_u(\xi)$ in blue. 
We observe that the blue solutions align with the red solutions almost perfectly, which shows that with a sufficient amount of data, an accurate feedback law can be learned. Note that the solutions are initialized through different initial conditions. Accordingly, solutions do not intersect or overlap, and instead are parallel.

\begin{remark}
The approximation of $(u_e^*(\xi),u_p^*(\xi))$ through $\psi_u(\xi)$ leads to almost indistinguishable performance. We point out that simply learning $\psi_V(\xi)$ to approximate $V(\xi)$ to subsequently define feedback laws through the formulas \eqref{eq:opt_feedback_law_e} and \eqref{eq:opt_feedback_law_p} while using the gradient $\nabla\psi_V(\xi)$ did not lead to satisfying results in numerical experiments.
\end{remark}

We observe that the feedback law of the evader is discontinuous on the (entire) $y$-axis and the feedback law of the pursuer is discontinuous on the positive $y$-axis. 
These discontinuities potentially lead to chattering in the implementation of the feedback laws subject to numerical errors or perturbations of the state (e.g., due to noise, sampling, or quantization).
Moreover, Fig. \ref{fig:sol_comparison} highlights the existence of a dispersal surface and a universal surface on the $y$-axis (see \cite[Sec. 9.4 \& 9.5.2]{Lewin2012}).
In the next section, we investigate the impact of sample-and-hold implementations of the feedback law $\psi_u$.

\section{Implications of sample-and-hold controller implementations} \label{sec:sample_and_hold}

In this section, we study the impact of sample-and-hold controller implementations, which is related to the Friedman theory in differential games as outlined in \cite[Ch. 8, Sec. 3.2]{bardi1997optimal}. 

\subsection{Sample-and-hold controller implementations} \label{sec:sub_sample}

Consider the closed-loop sample-and-hold dynamics
\begin{align}
   \dot{\xi}(t) = f(\xi,\bar{\psi}_{u,1}(t),\bar{\psi}_{u,2}(t)), \qquad \xi(0)=\xi_0 \label{eq:sample_hold}
\end{align}
where the inputs satisfy
\begin{align}
\begin{split}
    \bar{\psi}_{u,1}(t)&=\psi_{u,1}(\xi(\delta_e k)) \quad \forall t\in [\delta_e k, \delta_e (k+1)), \\
    \bar{\psi}_{u,2}(t)&=\psi_{u,2}(\xi(\delta_p k)) \quad \forall t\in [\delta_p k, \delta_p (k+1)),
\end{split} \label{eq:psi_bar}
\end{align}
for all $k\in \N$ and for fixed sampling periods $\delta_e,\delta_p \in \R_{>0}$\footnote{Instead of sample-and-hold implementations one can also study solutions based on intermittent sensing, i.e., the feedback law can be updated continuously, but the other player's strategy is only known at discrete time steps.}.
This setup corresponds to the scenario where the evader and the pursuer can only update their input at discrete time steps.
We define the set-valued map representing the time to end the game with the sample-and-hold dynamics \eqref{eq:sample_hold}-\eqref{eq:psi_bar} as $V^{\delta_{e}}_{\delta_p}:\mathcal{S}\rightrightarrows \R_{\geq 0}$. Since \eqref{eq:set_valued_feedback_laws} is set-valued, and since non-unique inputs can be used over time intervals of length $\delta_e$ and $\delta_p$, respectively, $V^{\delta_{e}}_{\delta_p}(\cdot)$ is set-valued as well. 
This scenario is particularly interesting if the two players have different sampling periods $\delta_e$ and $\delta_p$ and if 
\eqref{eq:sample_hold} are initialized close to the $y$-axis where the feedback laws \eqref{eq:opt_feedback_law_e} and \eqref{eq:opt_feedback_law_p} are discontinuous, as shown in Fig. \ref{fig:u_e_i_p}.

\begin{figure}[t!]
    \centering
    \begin{overpic}[width = 0.75\columnwidth]{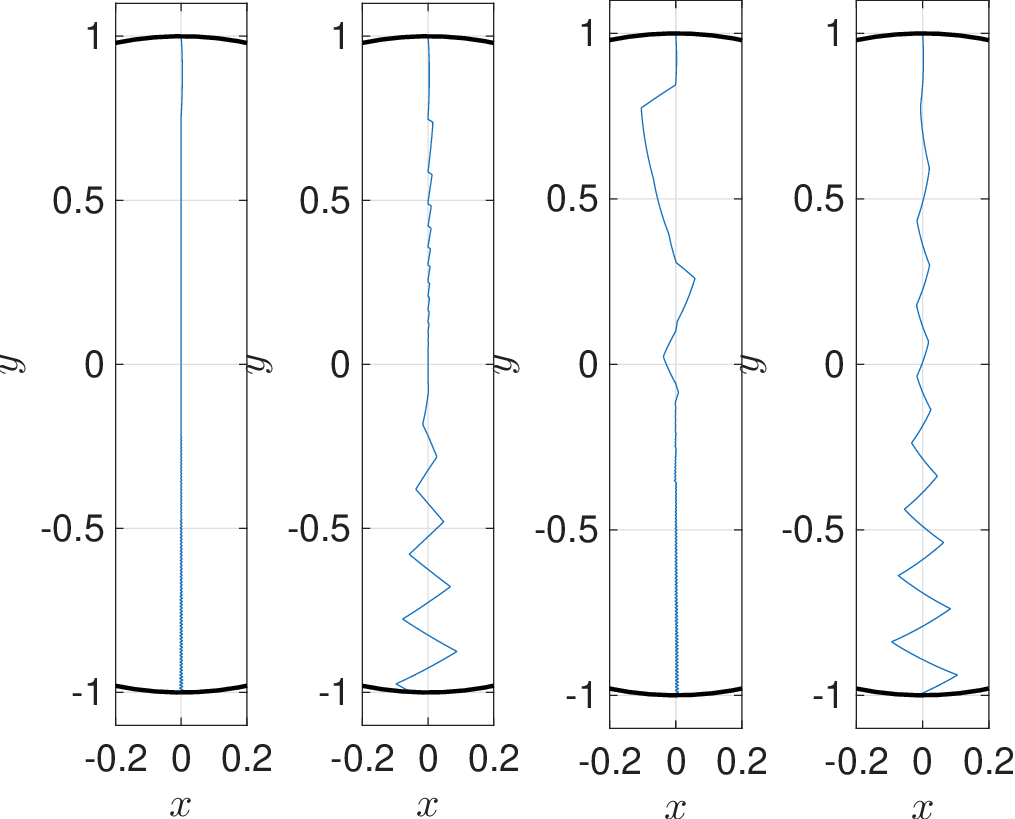}
    \end{overpic}
    \caption{Solutions starting at $\xi_0=[0,1]^\top$ using the sample-and-hold feedback  \eqref{eq:sample_hold}-\eqref{eq:psi_bar} for different sampling periods. From left to right, the sampling periods \eqref{eq:ex_sampling_rates} are used. The games end after time units in \eqref{eq:value_fun_sample}.}
    \label{fig:sample_and_hold}
\end{figure}

Fig. \ref{fig:sample_and_hold} shows closed-loop sample-and-hold implementations of the neural network based feedback law for 
$\xi_0=[0,1]^\top$. 
From left to right, the sampling periods  
\begin{align}
 \left[\begin{array}{c}
      \delta_e  \\
      \delta_p 
 \end{array} \right] \in \left\{ 
 \left[\begin{array}{c}
      0.01  \\
      0.01 
 \end{array}\right],
  \left[\begin{array}{c}
      0.2  \\
      0.01 
 \end{array}\right],
  \left[\begin{array}{c}
      0.01  \\
      0.2 
 \end{array}\right],   \left[\begin{array}{c}
      0.2  \\
      0.2 
 \end{array}\right]
 \right\} \label{eq:ex_sampling_rates}
\end{align}
are used.
The times to terminate the game are given by
\begin{align}
\begin{split}
    V^{0.01}_{0.01}(\xi_0) &= 3.4435, \qquad 
    V^{0.2}_{0.01}(\xi_0) = 3.4410, \\
    V^{0.01}_{0.2}(\xi_0) &= 3.2275,  \qquad
        V^{0.2}_{0.2}(\xi_0) = 3.3090.
    \end{split} \label{eq:value_fun_sample}
\end{align}
For small sampling periods (Fig. \ref{fig:sample_and_hold}, left), the solutions slide along the $y$-axis and the optimal solution corresponding to the continuous time-feedback law is recovered.
For $\delta_e$ large (second graph in Fig. \ref{fig:sample_and_hold}), the solution deviates dramatically from the continuous update of the feedback for $y<0$ (i.e., on the universal surface). However, the large deviations have little impact on the value of the value function. 
For $\delta_p$ large (third graph in Fig. \ref{fig:sample_and_hold}), the solution deviates dramatically from the continuous feedback law update for $y>0$ (i.e., on the dispersal surface). Here, the evader can benefit by terminating the game earlier. For the case where both $\delta_e$ and $\delta_p$ are large (Fig.~\ref{fig:sample_and_hold}, right), the solution behaves similarly to the first case. This suggests the possibility that the solution is more sensitive to $\delta_e$ than $\delta_p$.

\subsection{The rate of loss/gain}

In the previous subsection, we have seen that, depending on the sampling periods, individual players can obtain a significant advantage/disadvantage by using a sample-and-hold feedback 
implementation instead of a continuous-time implementation of the feedback law. 
We now further investigate how much a player can potentially gain or lose if they or their opponents use a sample-and-hold feedback strategy with a particular sampling period.
To this end, for $\delta>0$, we define the optimal control problem
\begin{align}
\begin{split}
    V_{\min}^\delta(\xi_0) &= \min_{\substack{u_e\in [-1,1]\\ u_p\in u_p^*(\xi_0)}} \int_0^\delta 1 \ dt + V(\xi(\delta)) \\
    \text{subject to } \quad & \dot{\xi}(t)=f(\xi(t),u_e,u_p), \qquad \xi(0)=\xi_0, \label{eq:optVdelMin}
\end{split}
\end{align}
optimizing a constant input of the evader over the time interval from $0$ to $\delta$, under the assumption that the pursuer chooses a constant optimal input depending on the initial condition $\xi_0$. 
In particular, under the assumption that the pursuer is selecting an arbitrary sample-and-hold optimal feedback selection for a sampling period $\delta_p = \delta$, \eqref{eq:optVdelMin} approximates how much the pursuer can lose (i.e., terminating the game earlier) if the evader uses a corresponding constant input over the time interval $[0,\delta]$.
We note that $V_{\min}^\delta(\cdot)$ in \eqref{eq:optVdelMin} can be simplified to
\begin{align}
    V_{\min}^\delta(\xi_0) &= \delta + \min_{\substack{u_e\in [-1,1]\\ u_p\in u_p^*(\xi_0)}}  V(\xi(\delta)). 
\end{align}
Moreover, $V(\xi(\delta))$ is not necessarily defined for all $\xi(\delta)$ and thus we define
\begin{align}
    V(\xi)= 0 \qquad \forall \ \xi  \in \R^2 \backslash \mathcal{S} \label{eq:V_extension}
\end{align}
so that $V_{\min}^\delta(\xi_0)$ is also well defined for initial conditions $\xi_0$, whose corresponding game of degree can end within $\delta$ time units.

Analogous to the definition \eqref{eq:optVdelMin} we define the corresponding loss/gain function from the perspective of the evader, i.e., we consider 
\begin{align}
\begin{split}
    V_{\max}^\delta(\xi_0) &= \delta + \max_{\substack{u_e\in u_e^*(\xi_0)\\ u_p\in \R}}  V(\xi(\delta)) \\
       \text{subject to } \quad & \dot{\xi}=f(\xi,u_e,u_p), \qquad \xi(0)=\xi_0.
\end{split} \label{eq:optVdelMax}
\end{align}
approximating the time that the evader might lose by choosing sample-and-hold feedback with $\delta_e = \delta$. 

\begin{remark}
Note that $V_{\min}^\delta(\cdot)$ and $V_{\max}^\delta(\cdot)$ are only approximations of the maximal gain/loss since they rely on the assumption that the opponent is using a constant input on the time interval $[0,\delta]$, which is in general restrictive.
Moreover, 
$V_{\min}^\delta(\cdot)$ and $V_{\max}^\delta(\cdot)$ only take the error from one sampling interval into account since the remainder of the game is approximated by $V(\xi(\delta))$. This is aligned with the setup discussed in \citep[Def. 2]{9661291}.
\end{remark}

\begin{figure}[t!]
    \centering
    \begin{overpic}[width = 0.48\columnwidth]{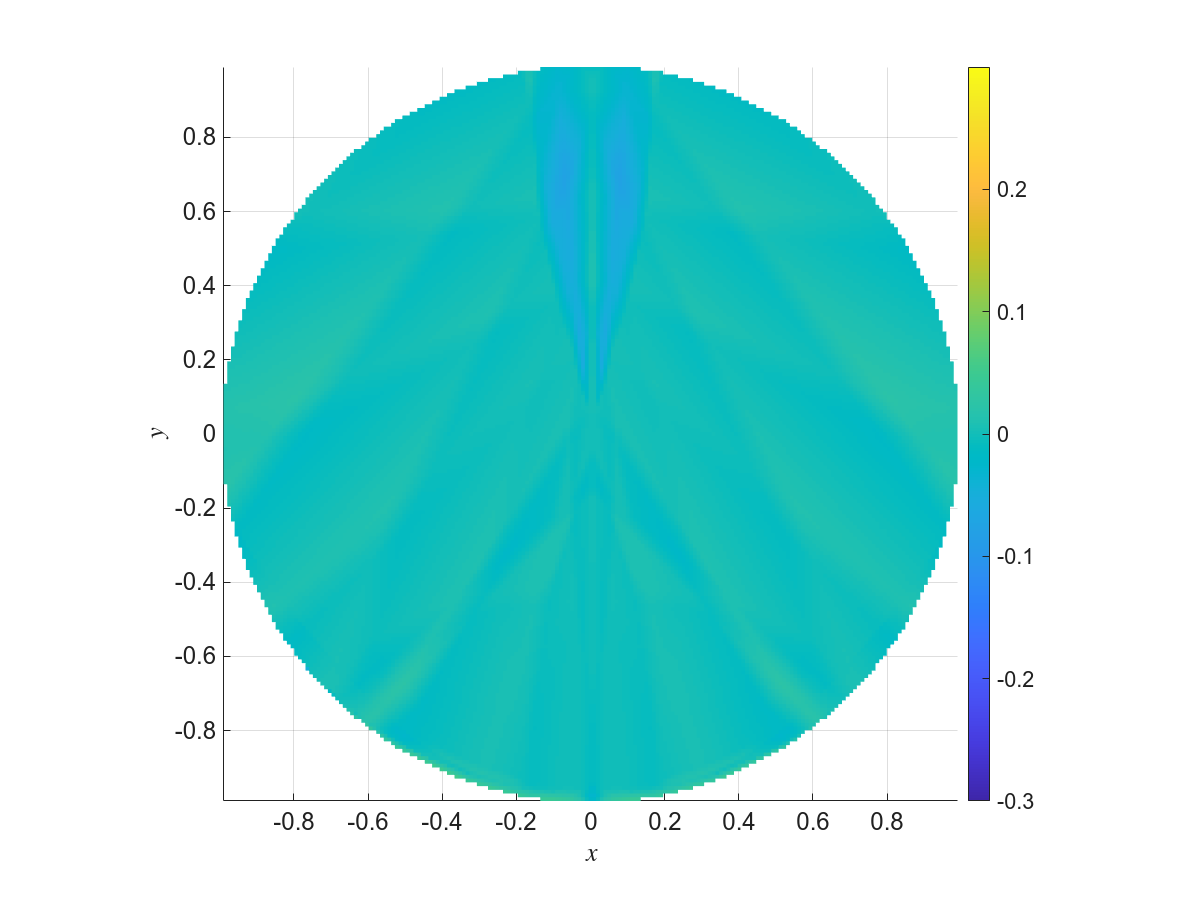}
    \end{overpic}
    \begin{overpic}[width = 0.48\columnwidth]{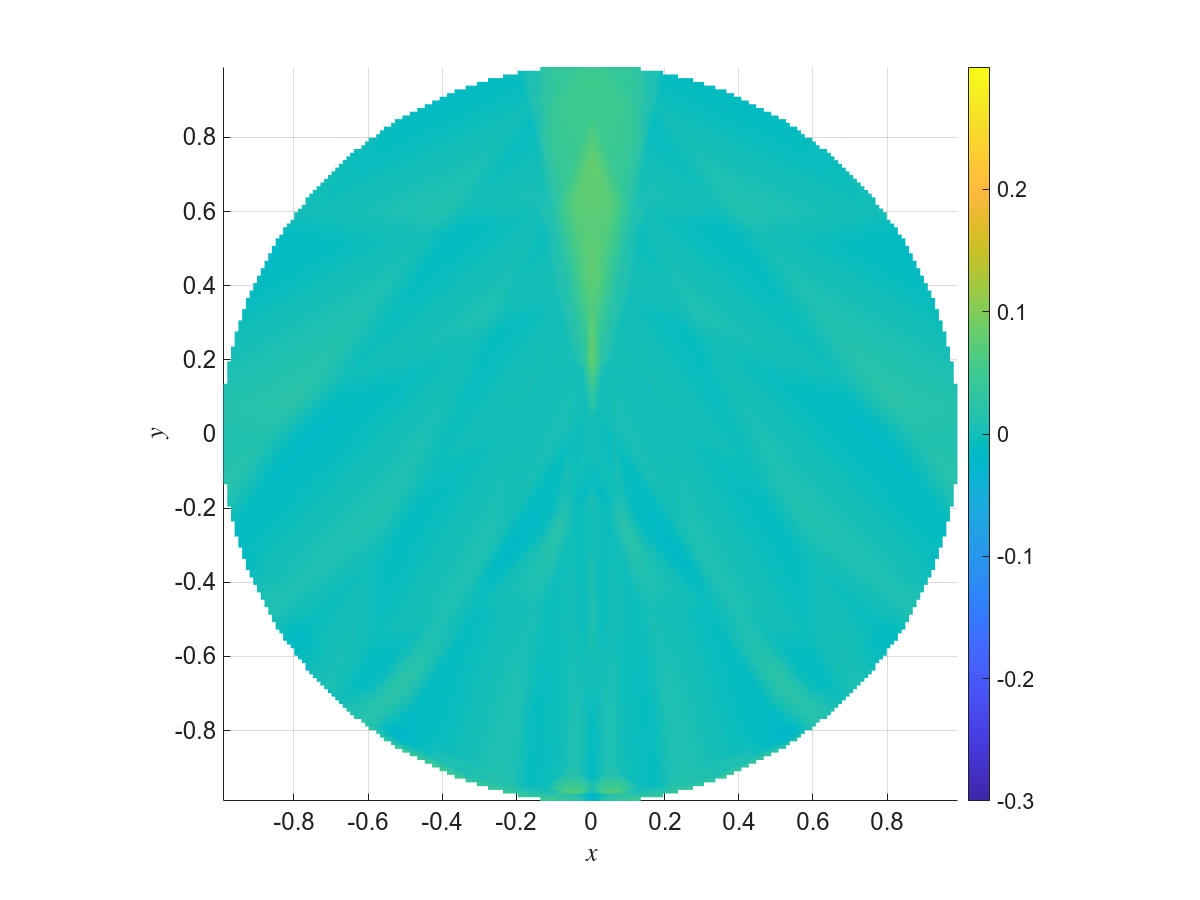}
    \end{overpic}

    \begin{overpic}[width = 0.48\columnwidth]{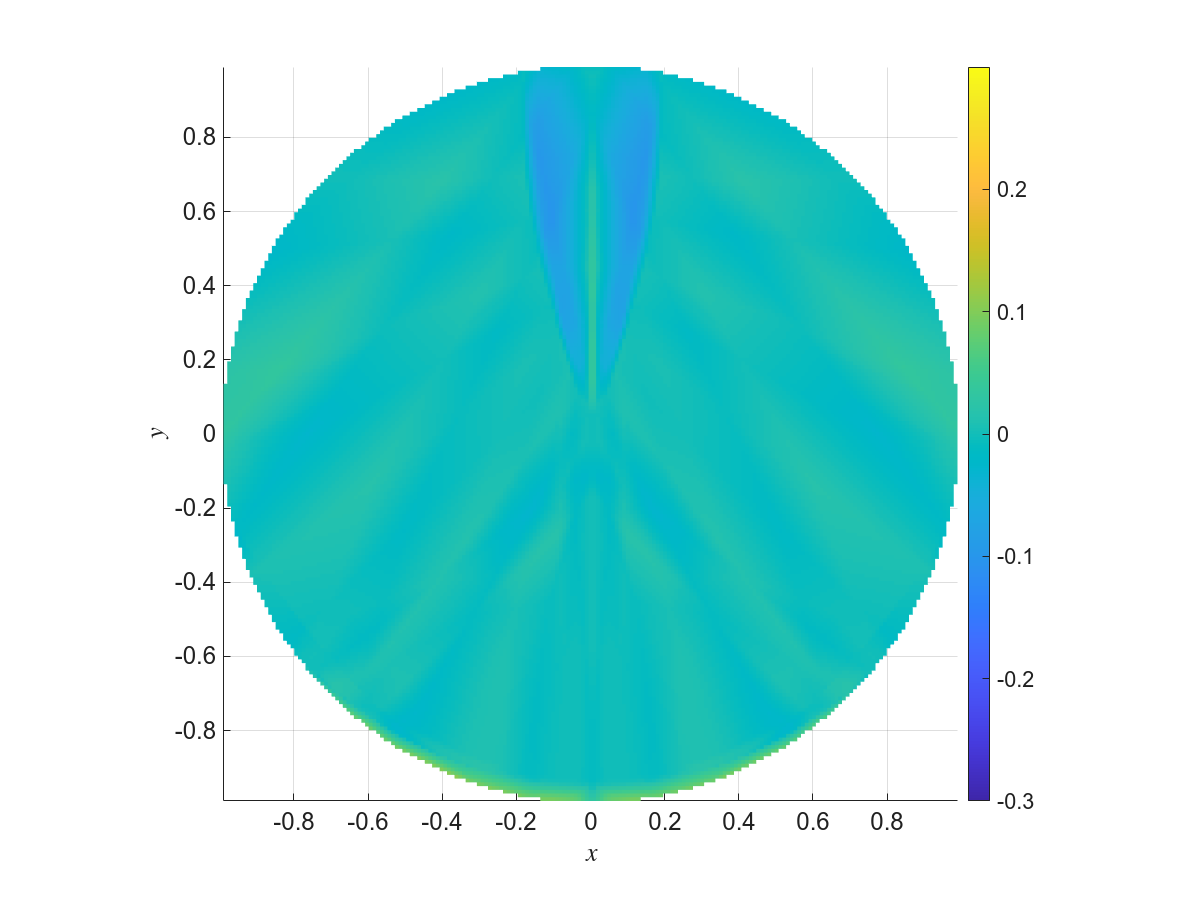}
    \end{overpic}
    \begin{overpic}[width = 0.48\columnwidth]{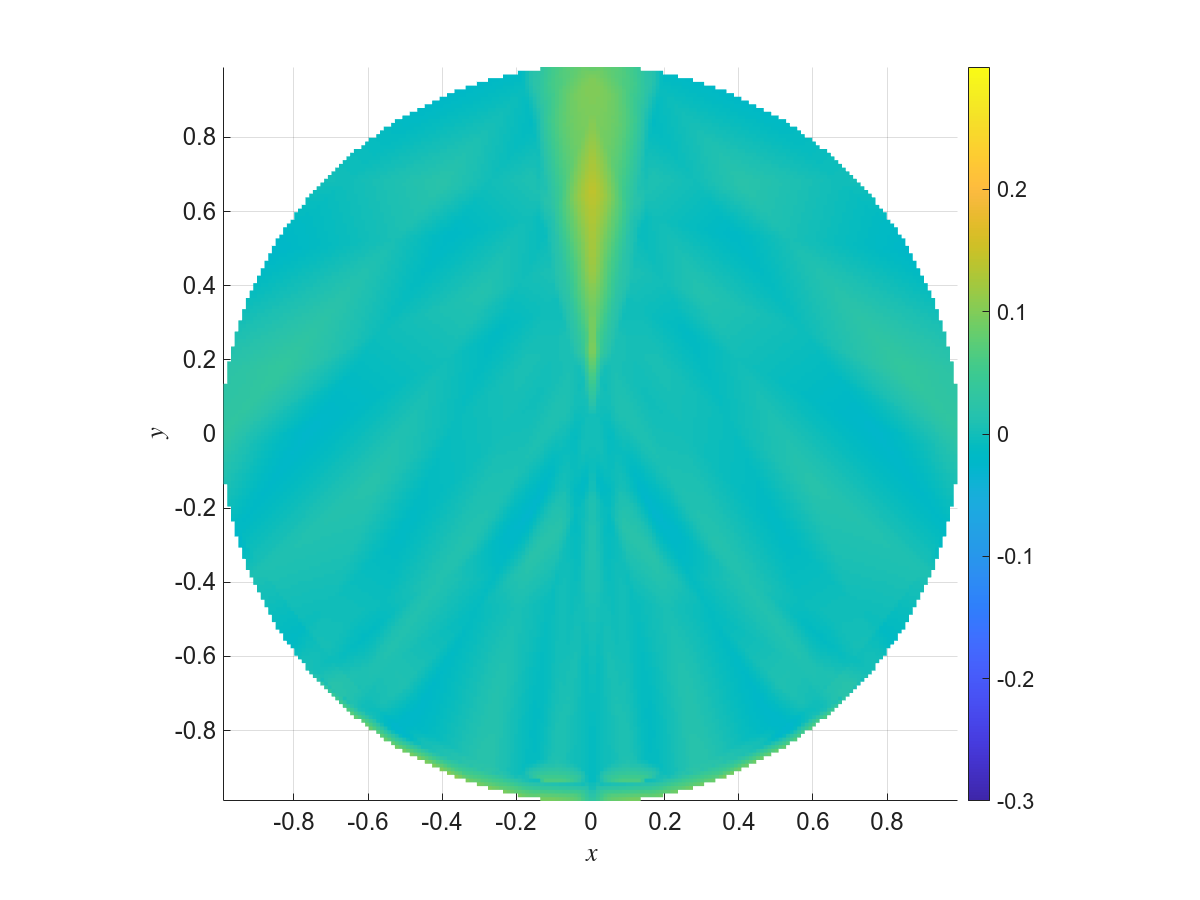}
    \end{overpic}

    \begin{overpic}[width = 0.48\columnwidth]{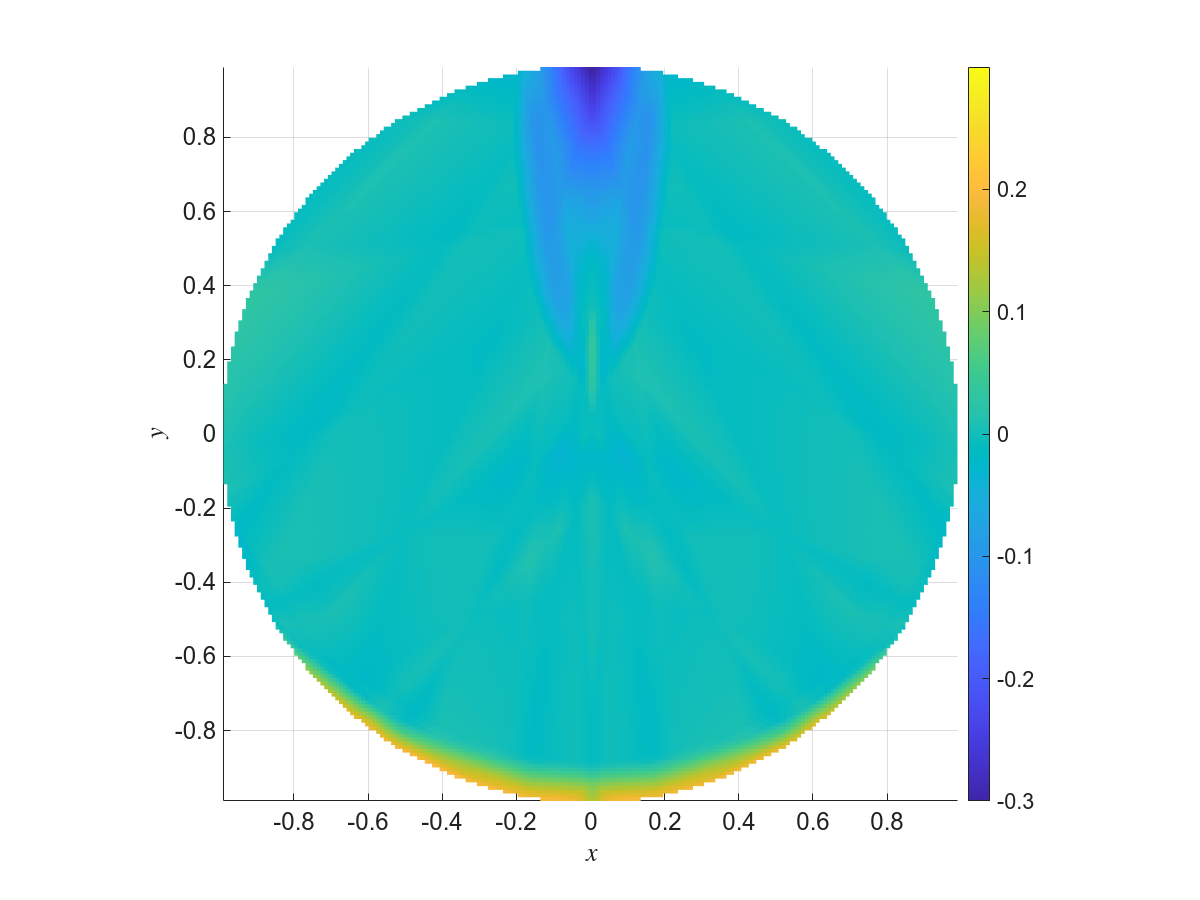}
    \end{overpic}
    \begin{overpic}[width = 0.48\columnwidth]{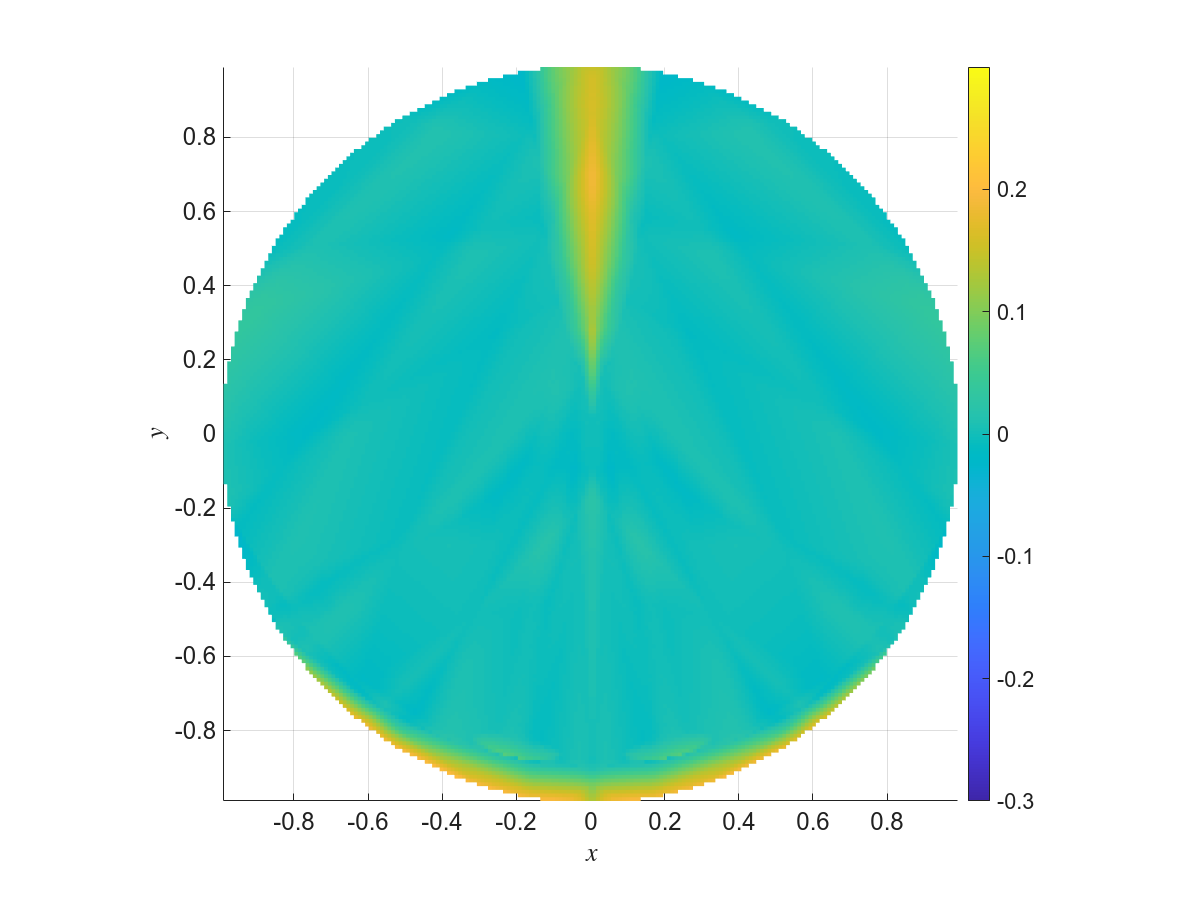}
    \end{overpic}
    \caption{Visualization of $V_{\min}^\delta(\cdot)$ (left) and $V_{\max}^\delta(\cdot)$ (right) for different parameter selections $\delta \in\{0.05,0.1,0.2\}$ (from top to bottom).}
    \label{fig:VV_max_min}
\end{figure}

In Fig. \ref{fig:VV_max_min}, the functions $V_{\min}^\delta(\cdot)$ (left) and $V_{\max}^\delta(\cdot)$ (right) are visualized for $\delta \in\{0.05,0.1,0.2\}$.
We observe that in a cone around the positive $y$-axis (around the dispersal surface) both players can benefit by exploiting the sample-and-hold controller implementation of their opponent. In particular, the evader can shorten the game (see Fig. \ref{fig:VV_max_min}, left) if the game is initialized close to the positive $y$-axis while the pursuer can prolong the game (see, Fig. \ref{fig:VV_max_min}, right). The amount of time that the game can be shortened/prolonged depends on the sampling period $\delta$. Consistent with the observation in Section \ref{sec:sub_sample} in Fig. \ref{fig:sample_and_hold}, the players cannot benefit on the negative $y$-axis (i.e., close to the universal surface) from the sample-hold-controller design of their opponent.
The increased values of $V_{\min}^\delta(\cdot)$ and $V_{\max}^\delta(\cdot)$ (compared with $V(\cdot)$) on the boundary of the game set $\mathcal{S}$ for $y<0$ are an artifact due to the extension of $V(\cdot)$ in \eqref{eq:V_extension} outside the game set.
The definition relies a piecewise constant input signal for $t\in[0,\delta]$. Thus, the optimal value functions $V_{\min}^\delta(\xi_0)$ and $V_{\max}^\delta(\xi_0)$ may differ from $V(\xi_0)$ even in the case where the optimal feedback laws are single valued. 
While this paper only investigates the impact of sample-and-hold controller designs for the particular case of the prying-pedestrian game, it opens up general questions for the analytical investigation of feedback designs in other pursuit-evasion and surveillance-evasion games.

\begin{remark}
    The evader's input \eqref{eq:opt_feedback_law_e} is piecewise constant and not affected by sampling if the $y$-axis is not crossed. The pursuer's input \eqref{eq:opt_feedback_law_p} on the other hand is not piecewise constant and thus may be affected by sampling everywhere in $\mathcal{S}$. This implies that the evader is robust to sampling while far away from the $y$-axis since the piecewise-constant structure of its control is derived from the optimal control law, rather than imposed by the sampling.
\end{remark}

\section{Conclusions} \label{sec:conclusions}

In this paper we have extended the results in \cite{prying_pedestrian} by showing that for the particular example of the prying-pedestrian surveillance-evasion differential game, neural networks can be used to obtain feedback laws from open-loop solution data. In addition, through numerical simulations, we have shown how individual players can benefit from sample-and-hold implementations of the feedback laws of their opponent close to the dispersal surface, while sample-and-hold implementations close to the universal surface of the game have almost no impact on the value of the game. Future work will focus on alternative pursuit-evasion games and study the generality of the numerical results derived here.

\bibliography{ifacconf}             

\end{document}